\documentclass[journal=jacsat,manuscript=article]{achemso}

\usepackage{amsmath}
\usepackage{mathtools}
\usepackage{amssymb}
\usepackage{caption}
\usepackage{subcaption}

\usepackage{chemformula} 
\usepackage[T1]{fontenc} 

\author{Samya Sen}
\affiliation{Department of Materials Science \& Engineering, Stanford University, Stanford, CA 94305, USA}
\altaffiliation{S.S. and C.D. contributed equally to this work}
\author{Changxin Dong}
\affiliation{Department of Materials Science \& Engineering, Stanford University, Stanford, CA 94305, USA}
\altaffiliation{S.S. and C.D. contributed equally to this work}
\author{Carolyn K. Jons}
\affiliation{Department of Materials Science \& Engineering, Stanford University, Stanford, CA 94305, USA}
\author{Wencke Reineking}
\affiliation{Veterinary Service Center, Department of Comparative Medicine, Stanford University School of Medicine, Stanford CA 94305, USA}
\author{Alakesh Alakesh}
\affiliation{Department of Materials Science \& Engineering, Stanford University, Stanford, CA 94305, USA}
\author{Noah Eckman}
\affiliation{Department of Chemical Engineering, Stanford University, Stanford, CA 94305, USA}
\author{Ye Eun Song}
\affiliation{Department of Materials Science \& Engineering, Stanford University, Stanford, CA 94305, USA}
\author{Alexander N. Prossnitz}
\affiliation{Department of Materials Science \& Engineering, Stanford University, Stanford, CA 94305, USA}
\author{Eric A. Appel}
\email{eappel@stanford.edu}
\affiliation{Department of Materials Science \& Engineering, Stanford University, Stanford, CA 94305, USA}
\alsoaffiliation{Department of Bioengineering, Stanford University, Stanford, CA 94305, USA}
\alsoaffiliation{Stanford ChEM-H, Stanford University, Stanford, CA 94305, USA}
\alsoaffiliation{Institute for Immunity, Transplantation and Infection, Stanford University School of Medicine, Stanford, CA 94305, USA}
\alsoaffiliation{Department of Pediatrics - Endocrinology, Stanford University School of Medicine, \\Stanford, CA 94305, USA}
\alsoaffiliation{Woods Institute for the Environment, Stanford University, Stanford CA 94305, USA}

\title[Evolving hydrogels for tunable release]{Evolving transport properties of dynamic hydrogels enable self-tuning of short- and long-term cargo delivery}

\keywords{Hydrogels, Cargo diffusion, Drug delivery, Rheology}

\begin{document}

\vspace{1em}

\begin{abstract}

\noindent
Hydrogels are crosslinked polymer networks with high water content, widely employed in biomedical applications such as drug delivery, tissue engineering, and regenerative medicine. Injectable, depot-forming hydrogels enable sustained release of therapeutic agents by modulating macromolecular diffusion through dynamic polymer networks. However, achieving reliable control over release kinetics remains a challenge, as the injection process induces shear-mediated disruption of transient crosslinks, leading to an initial burst release that can cause local toxicity and compromise therapeutic efficacy. Here, we present a hydrogel formulation strategy designed to restore network structure post-injection through rapid reformation of dynamic crosslinks, enabling time-dependent regulation of diffusion properties. By tuning viscoelastic parameters, including stress relaxation time and network recovery rate, we reduced the extent of burst release without compromising sustained delivery. Using model protein cargo, we demonstrate in both \emph{in vitro} and \emph{in vivo} settings that hydrogels with faster crosslink reformation kinetics exhibit significantly lower early-phase release while maintaining long-term delivery comparable to unmodified formulations. These results establish a mechanistic framework for decoupling short- and long-term release behavior, offering a broadly applicable strategy for precise drug delivery in soft tissue environments.
	
\end{abstract}


\section{Introduction}
\begin{figure}[!t]
	\centering
	\includegraphics[width=\linewidth]{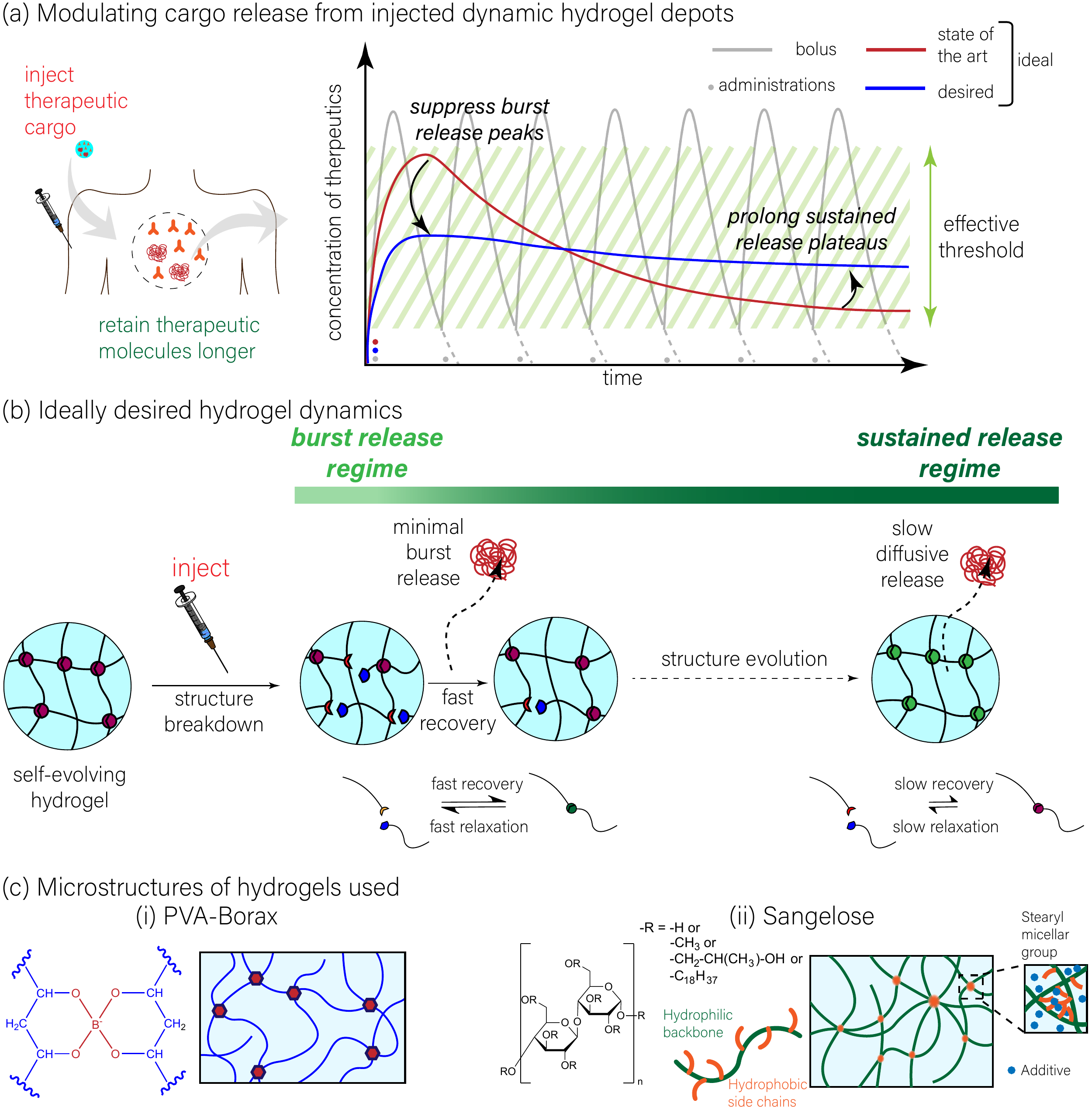}
	\caption{(a) Primary objective: modulate cargo release by designing the structural evolution of injectable hydrogels, minimizing burst release and extending sustained release. (b) crosslinking dynamics transition from fast to slow, altering cargo diffusion rates. Fast-relaxing networks reduce burst release, while slow-relaxing networks decrease diffusion coefficients, enhancing prolonged retention and sustained release. (c) Chemical structures and microstructures of the hydrogels: transiently crosslinked PVA-Borax networks formed \emph{via} covalent associative bonds between borate ions and PVA groups, and Sangelose hydrogels formed by hydrophobic clustering of stearyl groups on HPMC chains into dynamic, noncovalently crosslinked networks.}\label{fig:fig1}
\end{figure}

Hydrogels are crosslinked polymer networks with high water content, known for their biocompatibility and \emph{in vivo} tolerability. Their highly tunable mechanochemical properties make them versatile materials for diverse applications \cite{Correa2021,Appel2010,Mann2017,Appel_CSR2012,Webber2016,Li2016,Seiffert2015,Rodell2013,Buwalda2017,Mitragotri2014,Mooney2016,Dong2024,Lyla_TZP2025,Jons2025_BNAB}. In the biomedical field, hydrogels have garnered significant attention, leading to discoveries of novel properties and applications \cite{Correa2021,Mann2017,Webber2016,Li2016,Seiffert2015,Buwalda2017,Mitragotri2014,Mooney2016}. Injectable, depot-forming hydrogels represent a critical subfield, enabling minimally invasive, sustained drug delivery. Using the slower diffusion of macromolecules, including proteins and other biologics, from dynamic hydrogel networks, these systems achieve prolonged therapeutic release, often far exceeding the durations of intravenous administration (cf.~Fig.~\ref{fig:fig1}(a)), and elicit favorable immunological responses compared to bolus injections, underscoring their potential in advanced therapeutic applications \cite{Seiffert2015,Kasse2023,Grosskopf2019,Hernandez2018,Buwalda2017,Li2016,Mitragotri2014,Mann2017,Appel_CSR2012,Peppas2024,Mooney2016}.

A significant body of research focuses on developing hydrogels that prolong cargo retention in injected depots by tuning rheological properties (stiffness, relaxation time, yielding, and creep behavior) that control injection, depot persistence, and mass transport. The goal is to limit erosive cargo release and engineer a slow diffusive release rate \cite{Mann2017,Kasse2023,Mitragotri2014,Saouaf2021,Jons2022,Richbourg2021,Seiffert2015}. A typical sustained release profile from injectable hydrogel depots is shown in Fig.~\ref{fig:fig1}(a) (red). Physically crosslinked hydrogels with shear-thinning and self-healing properties enable easy injection and depot formation. Once injected, the release rates and cargo retention depend on the network structure, governed by its linear viscoelastic rheological properties \cite{Appel2010,Mann2017,Appel_CSR2012,Axpe,huang2024,Webber2016,Webber2022,Tibbitt2020,Tibbitt2021,Peppas2024,Mooney2016}. Increasing the stiffness of hydrogel networks (plateau shear modulus $G_0$) decreases the cargo release rate by reducing the effective mesh size ($\xi$), which impedes solute diffusion and lowers the diffusion coefficient compared to its Stokes-Einstein diffusivity ($\mathbb{D}/\mathbb{D}_0$). This relationship is supported by na\"ive network theory, which predicts $\xi \sim G_0^{-1/3}$, a trend confirmed by more advanced models \cite{RubinsteinColby,Offeddu2018,Axpe,Richbourg2021,Kasse2023,Saouaf2021,Tibbitt2020,Peppas2024}.

The focus has primarily been on prolonging the duration of release by retaining cargo within the depot, rather than tailoring release profiles to control both short-term burst release and sustained release over longer periods \cite{Seiffert2015,Li2016,Mann2017,Mitragotri2014,Peppas2024,Correa2021,Mooney2016}. Despite excellent sustained release characteristics, most injectable hydrogel platforms exhibit a phenomenon known as burst release, where a large fraction of encapsulated cargo is released within the first few hours following injection, often prior to full network recovery. In polymeric depots, burst release can account for 20-60\% of total cargo release within the first 4-8 hours, depending on hydrogel chemistry and injection site \cite{Chatterjee2019, Hu2021, Tibbitt2020}. Such uncontrolled early-phase release can lead to therapeutically suboptimal pharmacokinetics, including high peak plasma concentrations and short residence times, and has been associated with local tissue toxicity, especially in the case of potent biologics or immunomodulators \cite{Chatterjee2019, Mitragotri2014, Grosskopf2019}. Furthermore, burst release shortens the period over which therapeutic levels are maintained, limiting the overall efficacy of depot systems designed for extended delivery. Here, we define burst release as the fraction of total cargo released within the first 24 hours post-injection, a metric that is stricter yet easier to quantify, especially for \emph{in vivo} studies, compared with prior studies of shear-thinning or physically crosslinked hydrogels \cite{Webber2022, Tibbitt2021, Peppas2024}. The key challenge lies in minimizing this initial release without sacrificing the long-term, diffusion-governed sustained release phase. Decoupling these two temporal regimes to enable independent tuning of early and late release kinetics remains a significant barrier in the design of injectable depots.

Our observations indicate that hydrogel network dynamics, particularly the lifetimes and kinetics of reversible crosslinks that regulate network self-diffusion, play a critical role in controlling cargo transport, beyond the contributions of network topology (e.g., average mesh size related to stiffness) (cf.~Fig.~\ref{fig:fig2}). Although this effect has been noted in prior studies \cite{Tibbitt2020,Tibbitt2021,Webber2022}, it remains insufficiently characterized. We propose that the recovery of hydrogel structure following injection, during which crosslinks dissociate under high strain in confined flow, correlates with network relaxation behavior. In transient networks with similar stiffness $G_0$, which corresponds to similar equilibrium constants $K_{\rm eq}$, rapid stress relaxation (characterized by relaxation time $\tau_{\rm R} \simeq k_{\rm d}^{-1}$) implies faster structural recovery through crosslink reformation, with recovery time hypothesized to scale inversely with the association rate constant $k_{\rm a}$. Since $K_{\rm eq} = k_{\rm a} / k_{\rm d}$, maintaining similar stiffness across materials necessitates coordinated variation of both rate constants \cite{RubinsteinColby,Webber2022,Tibbitt2020,Webber2016,Seiffert2015,Tibbitt2021,Mann2017}. Hydrogels with slower crosslink kinetics exhibit reduced cargo diffusion under steady-state conditions due to their more static, fully recovered structure, favoring sustained release. However, their slow post-injection recovery enables excessive burst release, as loosely connected networks fail to contain rapid diffusive or advective transport under steep concentration gradients encountered in cargo-depleted environments (e.g., intramuscular or subcutaneous spaces). Conversely, hydrogels with faster crosslink kinetics restore network integrity more quickly after injection, suppressing burst release by more effectively resisting solute transport. Yet, these materials often display higher steady-state diffusion coefficients, diminishing their ability to sustain long-term release compared to slower-relaxing systems.

Achieving control over both short- and long-term cargo release profiles is thus challenging, often resulting in a compromise between minimizing burst release and ensuring prolonged release. In this work, we propose a strategy to address this challenge, as shown in Fig.~\ref{fig:fig1}(b). To reduce burst release, networks with fast crosslink kinetics are used initially for the injections. Subsequently, the network is allowed to evolve so the kinetics are slowed to achieve sustained release profiles similar to those of kinetically slow networks. The ideal release profile is shown in Fig.~\ref{fig:fig1}(a) (blue). Therefore, designing a hydrogel system that evolves its structure post-injection, transitioning from fast to slow crosslinking kinetics throughout the release period, is crucial.


\section{Results}

\subsection{\label{subsec:results-rheology}Hydrogels with Modular Transport Properties: Rheology and Diffusivity}
\begin{figure}[!t]
	\centering
	\includegraphics[width=\linewidth]{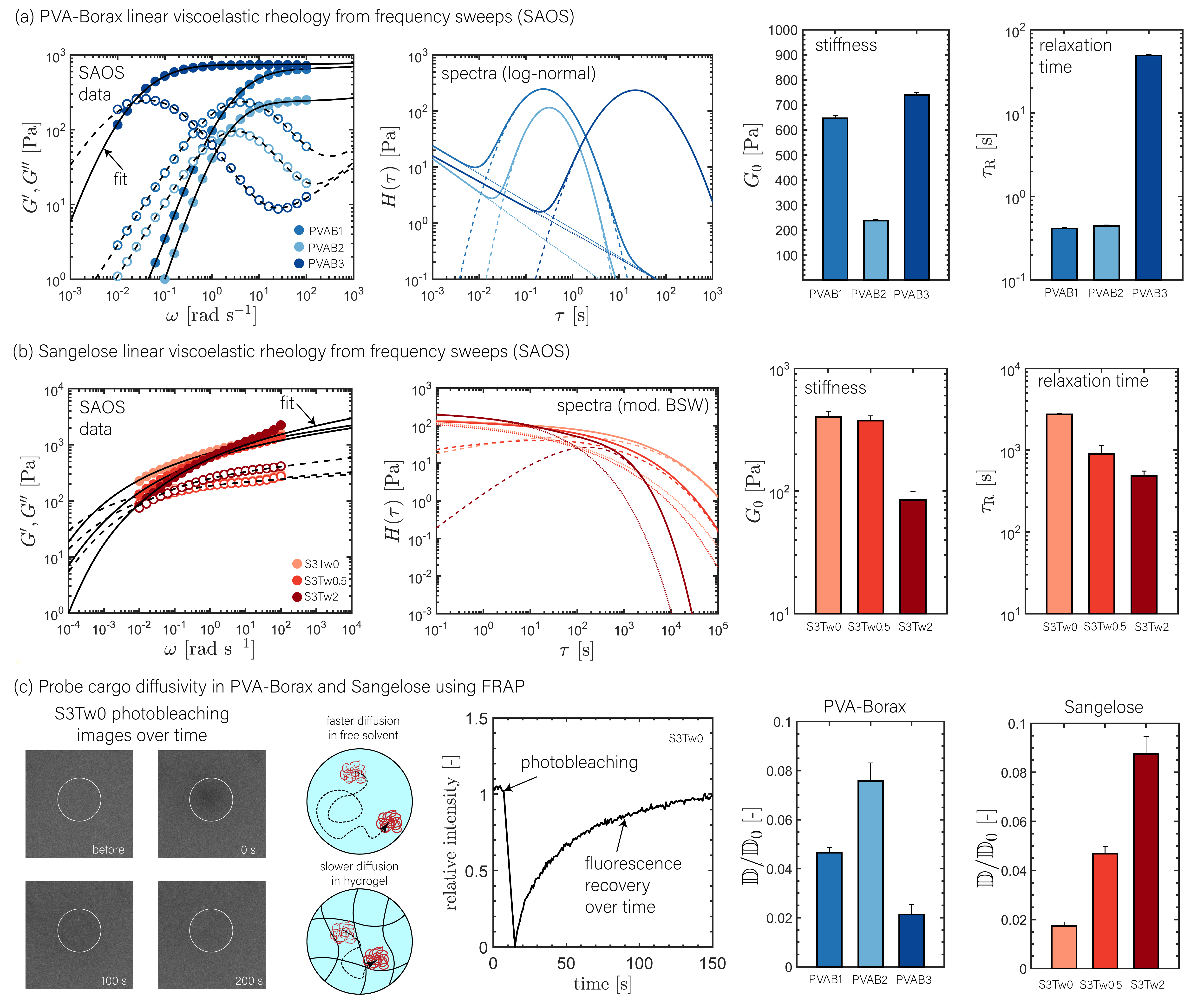}
	\caption{Transport properties of the hydrogels used. Frequency sweep linear viscoelastic rheology data for (a) PVA-Borax and (b) Sangelose hydrogels. First plot in each row: experimental and fitted linear viscoelastic moduli ($G^\prime$: filled circles, solid line; $G^{\prime\prime}$: open circles, dashed line) using log-normal (PVA-Borax) or modified BSW (Sangelose) relaxation spectra. Second plot: corresponding spectra with total (solid), viscoelastic/flow (dashed), and glassy (dotted) components. Bar plots show stiffness ($G_0$) and relaxation time ($\tau_{\rm R}$), highlighting hydrogel rheological tunability. (c) Cargo diffusivity measured by fluorescence recovery after photobleaching. Grayscale images show photobleaching and recovery, with exponential fits providing diffusion coefficients ($\mathbb{D}$) normalized to the free-solvent value ($\mathbb{D}_0$). Data reveals correlations between diffusivity, stiffness, and relaxation dynamics. Replicate measurements were taken for all data; $n = 3$ for rheology, and $n = 5$ for diffusivity; histograms show means with standard errors on the mean.}\label{fig:fig2}
\end{figure}

The timescale of network relaxation, directly related to crosslinking kinetics, provides an additional degree of freedom in tuning cargo diffusivity within hydrogel meshes \cite{Webber2022,Appel2014}. For networks with similar stiffness, those with distinct stress relaxation times exhibit different solute diffusivity, as observed in the hydrogels used in this study (cf.~Fig.~\ref{fig:fig2}). This property is leveraged to design materials with rheological properties that evolve between fast and slow crosslink kinetics, controlling solute diffusion while maintaining similar stiffness. To achieve this, small molecules, such as specific ions or surfactants, are incorporated to modulate crosslinking and promote faster network relaxation. While this may enhance burst release, a slow diffusion of these molecules from the network allows the structure to gradually regain slower crosslinking kinetics, lowering the diffusion coefficient and prolonging cargo retention for sustained release.

To demonstrate that networks with similar stiffness but different relaxation times exhibit varying cargo diffusivity (Fig.~\ref{fig:fig2}(a, c)), we formulated a model transient hydrogel network (Fig.~\ref{fig:fig1}(c)). The system consisted of poly(vinyl alcohol) chains covalently crosslinked by borate ions (see SI for formulation details). Borate- and boronic ester-crosslinked hydrogels are associative networks, where crosslinking kinetics directly influence relaxation behavior \cite{Martinetti2018,Tibbitt2020,Tan_PolymChem2017}. The relaxation time and dissociation rate constant are related as $\tau_{\rm R} \simeq k_{\rm d}^{-1}$, as stress relaxation occurs \emph{via} crosslink dissociation and reformation. Slowing dissociation kinetics significantly delays network relaxation. Modulating $p$H above the boric acid $pK_{\rm a} \simeq 9$ alters the crosslink dissociation kinetics, slowing relaxation without changing network stiffness, which remains governed by the average molecular weight between crosslinks. This is due to the $p$H-induced shift in association ($k_{\rm a}$) and dissociation ($k_{\rm d}$) rate constants, maintaining a near-invariant equilibrium constant $K_{\rm eq} = k_{\rm a} / k_{\rm d}$, which is correlated with $G_0$ \cite{RubinsteinColby,Webber2022,Tibbitt2020,Tibbitt2021,Webber2022}.

This is demonstrated in the shear rheological data shown in Fig.~\ref{fig:fig2}(a), which show linear-regime frequency sweeps and the continuous viscoelastic relaxation spectra fits for each material, and the simpler, tractable network parameters of stiffness ($G_0$) and mean relaxation time ($\tau_{\rm R}$) derived from analyzing the spectra (see SI for details). We see that going from PVAB1 (2.75\%~w/w PVA + 1.25\%~w/w borax, $p{\rm H} \simeq 10$) to PVAB2 (2.75\%~w/w PVA + 0.25\%~w/w borax, $p{\rm H} \simeq 10$), $G_0$ decreases due to a smaller density of crosslinkers, while $\tau_{\rm R}$ remains similar due to the same polymer concentration, and similar crosslink exchange dynamics due to the same $p$H. More interestingly, going from PVAB1 to PVAB3 (2.75\%~w/w PVA + 1.25\%~w/w borax, $p{\rm H} \simeq 13$), $G_0$ remains largely the same owing to same polymer and crosslinker concentration, but $\tau_{\rm R}$ increases significantly due to slower crosslink exchange at higher $p$H. This largely orthogonal control over stiffness and relaxation time allows us to isolate the effect of network relaxation on cargo diffusion and formulate materials with comparable stiffness but different relaxation dynamics (cf.~Fig.~\ref{fig:fig2}(a)).

For extending the same hypothesis to biocompatible materials suitable for \emph{in vivo} studies (see Sec.~\ref{subsec:results-invivo}), modified cellulose derived-hydrogels were used, which have been a suitable class of biomaterials for various translational studies \cite{Andrea_CRM2023,Jons2025_BNAB,Grosskopf2019,Rachel_AdvHealthMater2025,Hernandez2018,Correa2021,Bailey2024,Meany2025}. In particular, we used Sangelose hydrogels, which consists of hydroxypropyl methylcellulose stearoxy ether (HPMC-\ch{C_18}) chains (Fig.~\ref{fig:fig1}(c)). As reported in the literature, upon dissolution in aqueous media, the micellization of the stearyl groups pendant from the HPMC backbone forms a robust physical hydrogel network. While the micelles serve as reversible,
physical crosslinks between the HPMC polymer chains, the crosslink exchange dynamics are slow on account of the long, saturated, and highly hydrophobic stearic
chains \cite{Terukina2023,Rachel_AdvHealthMater2025,Meany2025}. The dynamics of the crosslinks may be tuned using surfactants that introduce free volume into the micellar domains, which can be used to modulate the network relaxation times and stiffness (see SI for formulation details). Shear rheological data for Sangelose hydrogels is shown in Fig.~\ref{fig:fig2}(b), which show linear-regime frequency sweeps and the continuous viscoelastic relaxation spectra fits for each material, and network parameters of stiffness ($G_0$) and mean relaxation time ($\tau_{\rm R}$) derived from analyzing the spectra (see SI for details). Each material contained 3\%~w/w Sangelose in phosphate buffered saline (PBS), and the difference was the amount of surfactant added (Tween 80). We see that going from S3Tw0 (no Tween 80 added) to S3Tw0.5 (0.5\%~w/w Tween 80), $G_0$ decreases marginally, while $\tau_{\rm R}$ decreases significantly, owing to the introduction of free volume into the micellar domains that accelerates the dynamic, noncovalent crosslink exchanges. Adding more surfactant, however, decreases both $G_0$ and $\tau_{\rm R}$ in S3Tw2 (2\%~w/w Tween 80).

The level of orthogonality is thus diminished, possible reasons being the physical, and thus non-specific nature of crosslinks arising from a non-uniform distribution of side chains along the cellulose backbone and thus also in each micellar domain, broader distribution of the molecular weight between crosslinks, and preferential absorption of surfactant molecules into denser crosslinked clusters. These factors may also lead to more broader relaxation behavior with significant contribution from glassy dynamics, as opposed to the largely viscoelastic fluid-like relaxation behavior of PVA-Borax with minimal glassy components. Extricating the stiffness associated with the viscoelastic network from the combined viscoelastic + glassy moduli is thus challenging and is reliably done using continuous spectra fits as employed in this work, but can still be model-dependent, and may not truly reflect the macroscopic flow properties of these materials, although the micro-scale cargo diffusion is still governed by the viscoelastic network dynamics \cite{Martinetti2018,Rachel_AdvHealthMater2025,huang2024}.

The diffusivity of model probe cargo in each material was measured using fluorescence recovery after photobleaching (FRAP) experiments. For PVA-Borax networks, bovine serum albumin (BSA, $M_{\rm w} \simeq 68~{\rm kDa}, R_{\rm g} \approx 4~{\rm nm}$) was used as the model protein cargo. This was chosen as such since the radius of gyration of BSA is comparable to the reported effective mesh size of PVA-Borax networks with stiffness used in this work \cite{Martinetti2018}, and the diffusion of cargo molecules within the network cargo release mechanism is dominated by network self-diffusion, and thus governed by the topological constraints and relaxation dynamics of the network \cite{Axpe,Peppas2024,Offeddu2018,Richbourg2021,Saouaf2021}. We see in Fig.~\ref{fig:fig2}(c) that increasing either $G_0$ or $\tau_{\rm R}$ decreases the diffusivity of BSA relative to that in the solvent/buffer $(\mathbb{D}/\mathbb{D}_0)$. This demonstrates that it is feasible to maintain network stiffness within reasonable limits, to retain mechanical properties such as depot persistence or injectability, and modulate the release rate of cargo molecules \emph{via} different relaxation times. For the Sangelose networks, human immunoglobulin-1 (hIgG, $M_{\rm w} \simeq 150~{\rm kDa}, R_{\rm g} \approx 7.8~{\rm nm}$) was used as the model protein to investigate cargo diffusivity within the hydrogel, which is a therapeutically relevant biomolecule used in characterization and developmental studies \cite{Jons2025_BNAB}. From Fig.~\ref{fig:fig2}(c), we observe similar trends as with PVA-Borax: increasing either $G_0$ or $\tau_{\rm R}$ decreases the diffusivity of hIgG relative to that in the solvent/buffer.

\begin{figure}[!t]
	\centering
	\includegraphics[width=\linewidth]{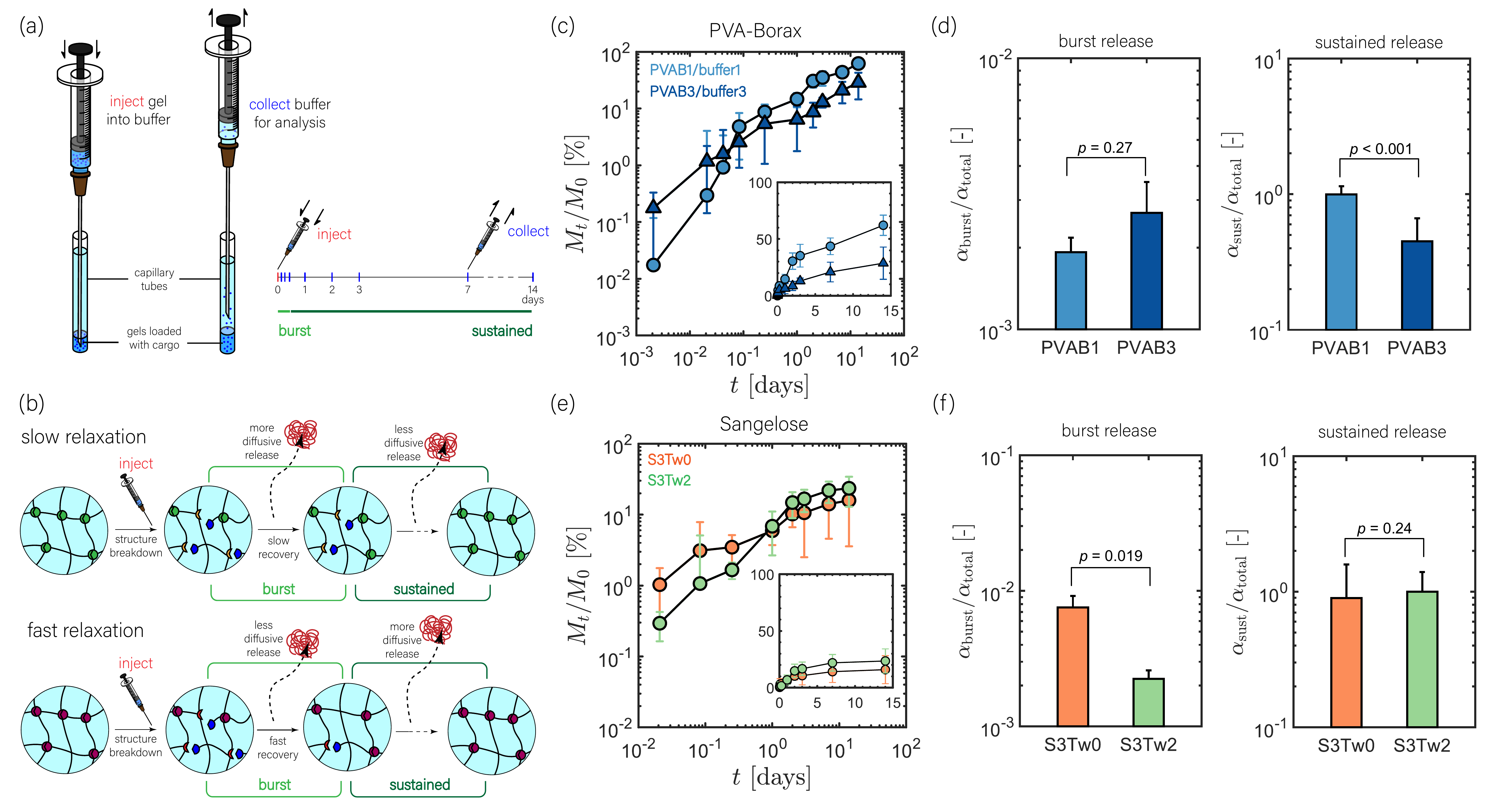}
	\caption{\emph{In vitro} cargo release studies. (a) Experimental setup: cargo-loaded hydrogel injected into buffer, with buffer collected over time to quantify protein release. (b) Hypothesized mechanisms for burst and sustained release in PVAB1/buffer1 and PVAB3/buffer3 systems. (c) Cumulative release profiles ($M_t/M_0$) for PVA-Borax hydrogels with BSA cargo in respective buffers, with inset showing linear-scale data. (d) Relative areas under burst and sustained release curves for PVA-Borax. (e) Cumulative release profiles ($M_t/M_0$) for Sangelose hydrogels with hIgG cargo in PBS, with linear-scale inset. (f) Relative areas under burst and sustained release curves for Sangelose. Replicate measurements were taken for all data; $n = 3$ for \emph{in vitro} release of both PVA-Borax and Sangelose; histograms show means with standard errors on the mean.}\label{fig:fig3}
\end{figure}

Based on these results, the effect of network topology and dynamics on cargo release from hydrogels was bench-marked using \emph{in vitro} release assays \cite{Kasse2023,Andrea_CRM2023,Lyla_TZP2025}. We injected two PVA-Borax hydrogels: PVAB1 with $G_0 \simeq 675.0~{\rm Pa}, \tau_{\rm R} \simeq 0.4~{\rm s}$, formulated in buffer1 with $p{\rm H} \simeq 10$, and PVAB3 with $G_0 \simeq 734.5~{\rm Pa}, \tau_{\rm R} \simeq 33.3~{\rm s}$, formulated in buffer3 with $p{\rm H} \simeq 13$, into capillary tubes with buffers of matching $p$H and hydroxyl ion concentrations (see SI for formulation details). Thus, these hydrogels had different relaxation times, but similar stiffness. This setup ensures a dynamic equilibrium between the hydrogel depot and the buffer interface, allowing only slow diffusive release of protein cargo. For release assays with PVA-Borax networks, BSA was used as the model protein cargo, for reasons explained previously. Cumulative release data ($M_t$) over 14 days, relative to the total cargo ($M_0$), is shown in Fig.\ref{fig:fig3}(c).

To quantify these differences, we calculated the fraction of release during the burst and sustained regimes using areas under the normalized release profiles (cf.~Fig.~\ref{fig:fig3}(d)), namely $\alpha_{\rm burst}/\alpha_{\rm tot}$ and $\alpha_{\rm sust}/\alpha_{\rm tot}$, where $\alpha_{T_1T_2}/\alpha_{\rm tot} = \sum_{n = T_1}^{T_2} (M_t)_n t_n$ is the normalized area between times $T_1$ and $T_2$. For burst release, $T_1 = 0$ and $T_2 = 1~{\rm day}$; for sustained release, $T_1 = 2~{\rm days}$ and $T_2 = 14~{\rm days}$. Since these areas are calculated for fractional mass release relative to total mass of cargo added to the hydrogel initially $(M_0)$, the fractions of burst and sustained release do not necessarily add to 100\%, indicating the presence of unreleased cargo in the depots which is typically the case for \emph{in vitro} assays. While \emph{in vitro} release assays rely primarily on diffusive cargo release in contrast with \emph{in vivo} studies on injected gel depots that release cargo molecules diffusively and erosively, \emph{in vitro} release data can be useful benchmarks for determining suitable candidate formulations for onward studies.

As hypothesized, PVAB1/buffer1 releases less cargo in the burst phase due to its faster network relaxation, but more cargo over longer durations compared to PVAB3/buffer3, whereas PVAB3/buffer3 has poorer burst release behavior. If PVAB1/buffer1 were functionalized to evolve into a slower relaxing network, we could potentially achieve both suppressed burst release and improved sustained release (\emph{vide infra}), which cannot be obtained using materials with static rheological properties.


\subsection{\label{subsec:results-release}\emph{In Vitro} Cargo Release of Evolving Hydrogels}
\begin{figure}[!t]
	\centering
	\includegraphics[width=\linewidth]{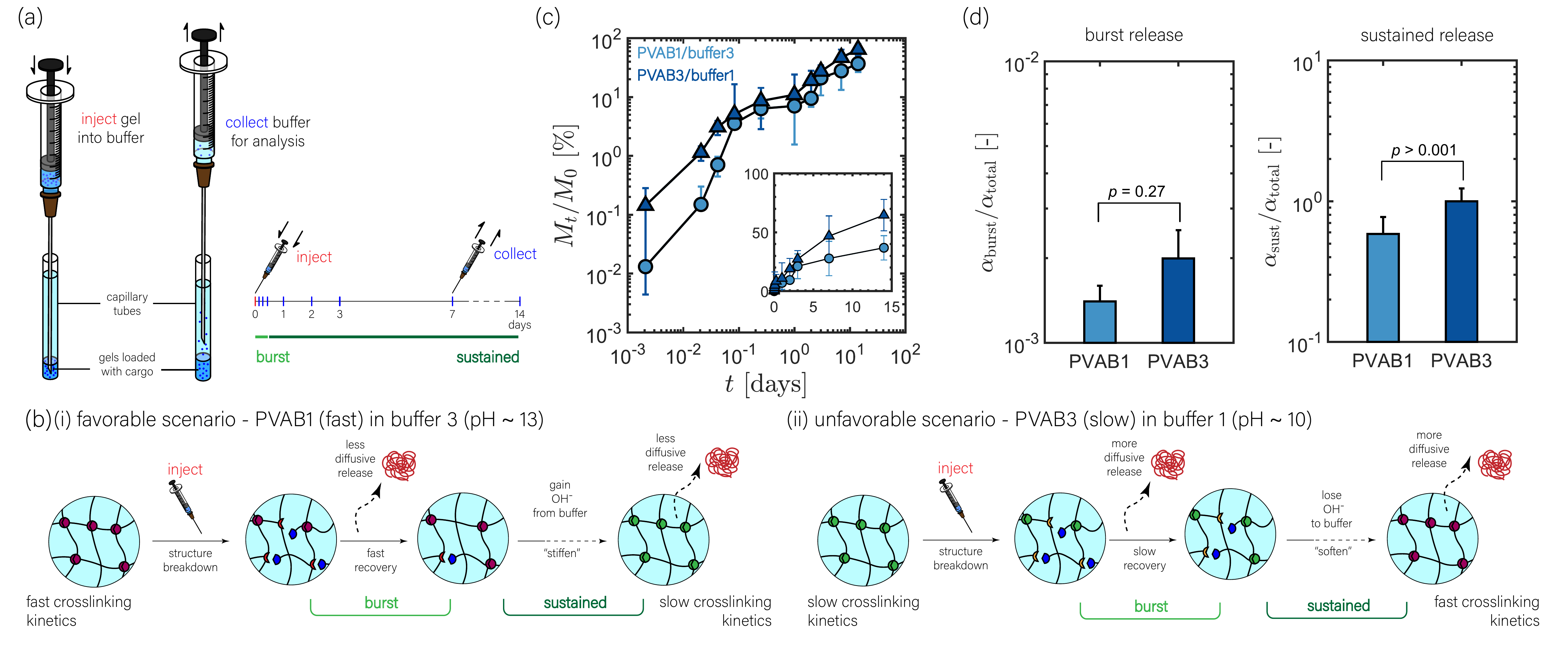}
	\caption{\emph{In vitro} cargo release studies with network evolution. (a) Experimental setup: cargo-loaded hydrogel injected into buffer, with buffer collected over time to quantify protein release. (b) Hypothesized mechanisms of burst and sustained release for PVAB1/buffer3 and PVAB3/buffer1 systems, illustrating network evolution leading to (i) favorable or (ii) unfavorable release. (c) Cumulative release profiles ($M_t/M_0$) for PVA-Borax hydrogels with BSA cargo in opposite buffers, with linear-scale inset. (d) Relative areas under burst and sustained release curves for PVA-Borax. Replicate measurements were taken for all data; $n = 3$ for \emph{in vitro} release of PVA-Borax; histograms show means with standard errors on the mean.}\label{fig:fig4}
\end{figure}

To test our hypothesis of modulating cargo release profiles through small molecule exchange, we injected the same PVA-Borax hydrogels into buffers with opposite $p$H conditions and monitored the cumulative cargo release \emph{in vitro} (cf.~Fig.~\ref{fig:fig4}(c)) and fractional cargo release data (cf.~Fig.~\ref{fig:fig4}(d)). PVAB3/buffer1, with slower crosslink exchange, released more cargo in the burst regime, but upon \ch{OH^-} exchange with the buffer, its pH decreased, accelerating crosslinking and cargo diffusion, leading to higher sustained release. In contrast, PVAB1/buffer3 released less cargo in the burst phase due to faster dynamics. However, as the structure evolved through \ch{OH^-} exchange, the network slowed, reducing diffusivity and sustaining a smaller release fraction. This illustrates the possibility for real-time modulation of cargo release from hydrogel depots, enabling the design of hybrid release profiles through controlled hydrogel and buffer interactions.

To translate the evolving network strategy for \emph{in vivo} applications, the design must accommodate physiological compatibility conditions, where \ch{H^+} or \ch{OH^-} ion exchange is not feasible and a $p{\rm H} \simeq 7$ is preferred for biocompatibility \cite{Mooney2016,Mitragotri2014,Correa2021,Buwalda2017}. We implemented evolving networks \emph{in vivo} using Sangelose hydrogels, which are biocompatible (Fig.~\ref{fig:fig1}(c)) \cite{Terukina2023,Rachel_AdvHealthMater2025,Meany2025}. Surfactants were used to modulate the crosslinking kinetics by altering hydrophobic side-chain interactions, producing materials with significantly different stiffness and different relaxation times (cf.~Fig.~\ref{fig:fig2}(b)). Similar \emph{in vitro} release assays as PVA-Borax were employed to test the hypothesis of release tuned evolving rheology in Sangelose networks. Two Sangelose formulations in phosphate buffered saline (PBS, $p{\rm H} \simeq 7.4$), S3Tw0 ($G_0 \simeq 402.9~{\rm Pa}, \tau_{\rm R} \simeq 2740.8~{\rm s}$) and S3Tw2 ($G_0 \simeq 84.4~{\rm Pa}, \tau_{\rm R} \simeq 483.3~{\rm s}$, 2\%~w/w Tween 80) were each injected into capillary tubes containing PBS, and the release of cargo was monitored over time (see SI for formulation details). As with FRAP experiments (Fig.~\ref{fig:fig2}(c)), hIgG was used as the model protein cargo. These materials exhibited distinct diffusion coefficients due to varying network relaxation dynamics (cf.~Fig.~\ref{fig:fig2}(c)). In \emph{in vitro} release assays, Sangelose hydrogels with shorter relaxation times released less hIgG cargo in the burst phase compared to those with longer relaxation times, while both exhibited similar sustained release profiles (Fig.~\ref{fig:fig3}(e) and (f)). This behavior is attributed to the fast-relaxing network gradually losing surfactant molecules to the buffer, evolving its rheological properties toward those of the slower-relaxing network, resulting in comparable release profiles over time. These results are therefore promising for further studies on the effect of evolving rheological and network properties in animal experiments (\emph{vide infra}).


\subsection{\label{subsec:results-invivo}\emph{In Vivo} Pharmacokinetics of Evolving Hydrogels with Therapeutic Cargo in Mice}
\begin{figure}[!t]
    \centering
    \includegraphics[width=\linewidth]{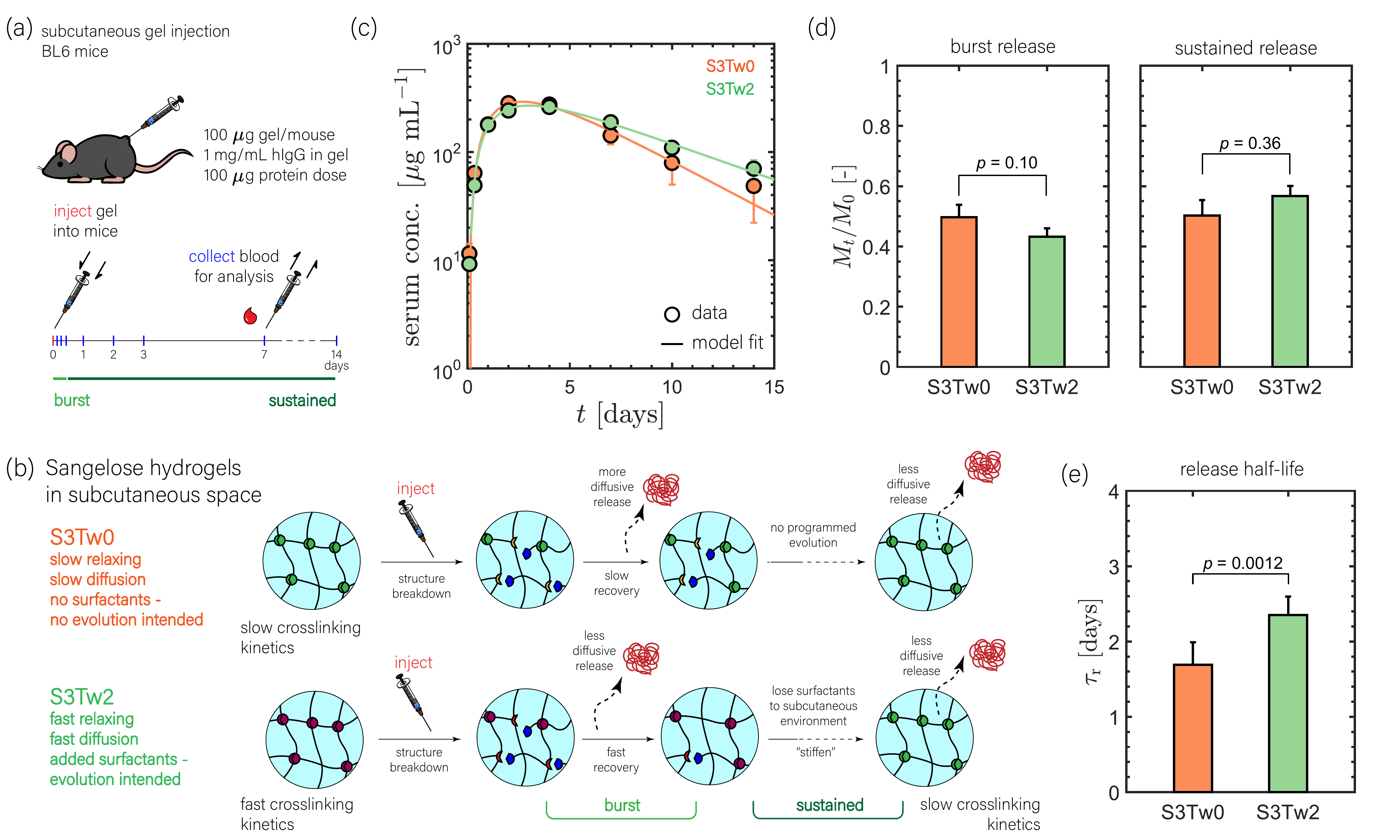}
    \caption{\emph{In vivo} pharmacokinetic analysis of Sangelose hydrogels loaded with human IgG cargo. (a) BL6 mice were injected subcutaneously with hydrogels containing hIgG, and serum levels were measured over time. (b) Schematic of proposed release mechanism: Tween 80-functionalized hydrogel (S3Tw2) evolves post-injection, enabling delayed onset and prolonged release. (c) Serum hIgG concentration profiles for S3Tw0 and S3Tw2 formulations; solid lines indicate two-compartment pharmacokinetic model fits. (d) Cumulative serum exposure over time shows extended delivery from S3Tw2. (e) Extracted release and elimination half-lives from model fits; S3Tw2 demonstrates a significantly longer release half-life, reflecting enhanced control over early-phase release kinetics. While neither formulation exhibits a large classical burst, the extended rise time and delayed peak of S3Tw2 indicate successful temporal modulation of cargo transport. Replicate measurements were taken for all data; $n = 5$ animals for \emph{in vivo} studies of both Sangelose formulations; histograms show means with standard errors on the mean.}\label{fig:fig5}
\end{figure}

In addition to biocompatibility, facile formulation, and sustained release capabilities, Sangelose hydrogels are also shear-thinning and recover their structure quickly after being subjected to high shear stresses during extrusion, which make them suitable candidates for use as injectable, depot-forming materials \cite{Rachel_AdvHealthMater2025,Andrea_CRM2023,Lyla_TZP2025,Meany2025}. To evaluate the translational relevance of our hydrogel design, we tested the pharmacokinetics (PK) of surfactant-modified Sangelose hydrogels \emph{in vivo} using a subcutaneous mouse model (Fig.~\ref{fig:fig5}). Human IgG (hIgG) was encapsulated and delivered \emph{via} two formulations, S3Tw0 (fast-relaxing) and S3Tw2 (slow-evolving), and serum protein levels were measured over time. Both groups exhibited typical PK curves, with an initial rise in serum concentration followed by gradual clearance. S3Tw0 reached a peak serum concentration ($c_{\rm max}$) of $282.27~\mu{\rm g~mL}^{-1}$ at 48~h ($t_{\rm max}$), while S3Tw2 peaked later at 96~h with a slightly lower $c_{\rm max}$ of $260.69~\mu{\rm g~mL}^{-1}$. While both profiles showed limited early-phase spikes in concentration, the later peak and slower ascent of S3Tw2 suggest attenuated early-phase release kinetics. However, the absence of a large, classical burst release in either system, often defined as $> 30\%$ of total cargo released within the first 12-24~h, highlights a limitation in modeling burst behavior strictly from in vivo serum levels. Instead, differences are more accurately interpreted through kinetic modulation of the release phase.

Notably, area under the curve (AUC) analysis revealed a larger total exposure for S3Tw2 (55,181.69 $\mu{\rm g~h~mL}^{-1}$) compared to S3Tw0 (50,752.97 $\mu{\rm g~h~mL}^{-1}$), indicating prolonged systemic presence. Two-compartment PK modeling was used to deconvolute cargo release half-life from the depot ($\tau_{\rm r}$) and elimination half-life ($\tau_{\rm e}$). S3Tw2 exhibited a significantly longer release half-life ($6.02 \pm 1.04~{\rm d}$) compared to S3Tw0 ($3.19 \pm 1.27~{\rm d}$), while elimination half-lives were comparable ($\sim 1~{\rm d}$), consistent with literature for similar protein therapeutics and administration routes \cite{Kasse2023,Mann_STM2023}.

Histological examination was also performed ex vivo on skin samples harvested from animals after 14~d post injection (see SI for details). No visible differences were observed between skin samples across groups containing Sangelose hydrogels with or without surfactants compared to both empty gels and those containing hIgG cargo, clearly demonstrating the \emph{in vivo} tolerance of these gels, indicating that the differences observed in the PK between S3Tw0 and S3Tw2 were due to difference in network properties of the gels; healthy animals with no treatment were used as the reference. These data suggest that modulating hydrogel crosslinking kinetics shifts the shape of the early-phase release curve, delaying $t_{\rm max}$ and extending the window of release. While overt burst release was not observed \emph{in vivo}, potentially due to the intrinsic physical barriers of the subcutaneous environment or the already favorable rheological properties of the baseline hydrogel, the difference in early release rates supports the utility of dynamic network design in fine-tuning pharmacokinetics. The findings reinforce the premise that controlling early-phase release kinetics, even without large bursts, is crucial for optimizing therapeutic delivery.


\section{Discussion}
This study demonstrates a strategy for engineering injectable dynamic hydrogels capable of evolving their transport properties post-injection to modulate early- and late-phase drug release kinetics. By tuning the crosslinking kinetics \emph{via} incorporation of surfactants that exchange with the physiological environment, we created hydrogels that rapidly recover structural integrity following injection and gradually transition to slower-relaxing networks over time. This approach delays cargo release onset and extends therapeutic availability.

While the \emph{in vivo} pharmacokinetics did not exhibit large, classical burst release, commonly defined as the rapid release of $> 30\%$ of total drug within hours, differences between formulations were evident in the timing and rate of release onset. The slower rise to peak serum concentration and significantly longer depot release half-life observed in S3Tw2 highlight the system’s ability to modulate early-phase kinetics without compromising long-term exposure. These outcomes underscore a more nuanced understanding of burst release as a continuum of early transport behaviors, where shaping the temporal profile, even in the absence of sharp initial spikes, can still yield important therapeutic benefits. This level of kinetic control is critical for applications where tight regulation of systemic exposure is essential. For example, reducing steep concentration gradients early post-injection can help avoid toxicity from immunomodulatory agents like GLP-1 agonists \cite{Andrea_CRM2023,Lyla_TZP2025}, while extending delivery is vital for resource-intensive therapeutics such as long-acting antiretrovirals and vaccines \cite{Kasse2023,Jons2025_BNAB}.

Our findings support a broader design framework for injectable depots that emphasizes \emph{temporal flexibility} over fixed rheological properties. Rather than focusing solely on minimizing burst release, future injectable materials can leverage \emph{adaptive crosslinking dynamics} to shape the full release curve, enabling application-specific tuning. This concept advances the field of adaptable soft materials and provides a platform for dynamic, context-responsive drug delivery. Future directions include incorporating biomolecular triggers or environmental sensors to achieve even finer control over network evolution and pharmacokinetic profiles in complex physiological settings.


\section{Methods}

\subsection{Materials\label{subsec:materials}}
PVA-Borax gels were prepared by mixing poly(vinyl alcohol) (PVA) solution (85-124 kDa, Sigma, 6\%~w/w stock) with sodium tetraborate (\ch{Na_2B_4O_7}) solution (Borax, Sigma, 6\%~w/w stock) in appropriate proportions to achieve 2.75-1.25\%~w/w or 2.75-0.25\%~w/w concentrations. The $p$H was adjusted with 1~M sodium hydroxide. Sangelose hydrogels were prepared by dissolving Sangelose 90L (Daido Chemical Co., Osaka, Japan) powder in phosphate-buffered saline (PBS) to form a 6\%~w/w stock, which was then diluted to 3\%~w/w. Polysorbate 80 (Tween 80, Sigma) surfactant was added during dilution as needed. Detailed formulation procedures are provided in the SI.

\subsection{Shear Rheometry}
Rheological measurements were conducted using a Discovery Hybrid Rheometer (DHR-2, TA Instruments). Details of rheometry procedures and continuous relaxation spectra analysis are provided in the SI.

\subsection{\emph{In Vitro} Diffusivity Measurements}
The diffusion coefficients of probe cargoes in the hydrogels were determined using fluorescence recovery after photobleaching (FRAP), procedure is detailed in the SI. Fluorescein isothiocyanate-tagged bovine serum albumin (BSA, Sigma, \(M_{\rm w} \simeq 68~{\rm kDa}\)) was used for PVA-Borax, and Alexa 647-tagged human immunoglobulin (hIgG, Sigma, \(M_{\rm w} \simeq 150~{\rm kDa}\)) for Sangelose. Briefly, fluorescent cargo was bleached with a high-intensity laser, and the fluorescence recovery time constant, \(t_{1/2}\), was used to calculate the diffusion coefficient \(\mathbb{D} = \gamma_\mathbb{D} (\rho^2 / t_{1/2})\), where \(\gamma_\mathbb{D}\) is an empirically determined constant \cite{Axelrod1976}.

\subsection{\emph{In Vitro} Release Assays}
Release assays were conducted by injecting $100~\mu{\rm L}$ of PVA-Borax and Sangelose hydrogels loaded with FITC-BSA and Alexa-hIgG respectively (concentration of $10~{\rm mg~mL}^{-1}$) into a capillary tube filled with $400~\mu{\rm L}$ of the appropriate buffer. At regular intervals, the supernatant was removed and analyzed to measure the amount of cargo present. Detailed procedures are described in the SI.

\subsection{\emph{In Vivo} Pharmacokinetics Studies}
Animal studies were conducted on female BL6 mice (6-8~weeks old) administered hIgG \emph{via} subcutaneous (SC) injection of 100~$\mu$L Sangelose hydrogel containing 1~mg hIgG/mouse. Serum samples were collected from the tail at 2 and 8~h, and 1, 2, 4, 7, 10, and 14~d (\(n = 4\)) for both groups (S3Tw0 and S3Tw2), and analyzed for hIgG concentration by ELISA. Details of gel preparation, injection, ELISA, and pharmacokinetics modeling are provided in the SI. NIH guidelines for the care and use of laboratory animals (NIH Publication \#85-23 Rev.~1985) have been observed. All animal studies were performed with the approval of Stanford Administrative Panel on Laboratory Animal Care.


\section{Acknowledgements}
The authors thank the Bill \& Melinda Gates Foundation (INV-008642) for their financial support of this work in the development of facile and effective injectable drug delivery platforms. The probe diffusivity data using FRAP were obtained using confocal microscopy at the Stanford Cell Sciences Imaging Facility (CSIF).


\section{Conflicts of Interest}
E.A.A., S.S., C.D., and Y.E.S. are listed as inventors on patent applications describing the hydrogel technology reported in this manuscript. E.A.A. is a co-founder, equity holder, and advisor to Appel Sauce Studios LLC, which holds a global exclusive license to these technologies from Stanford University. All other authors declare no competing financial interests.


\begin{suppinfo}
Supporting Information is available from the Wiley Online Library or from the author.
\end{suppinfo}

\bibliography{references}

\newpage

\begin{tocentry}
  \includegraphics[width=\linewidth]{TOC.png}
\end{tocentry}

\end{document}